\documentclass[sigconf]{acmart}

\renewcommand\footnotetextcopyrightpermission[1]{} 
\makeatletter
\renewcommand\@formatdoi[1]{\ignorespaces}
\makeatother

\usepackage{graphicx}
\usepackage{tikz}
\usepackage{balance}
\usepackage{cite}
\usepackage{breakcites}

\AtBeginDocument{%
  \providecommand\BibTeX{{%
    \normalfont B\kern-0.5em{\scshape i\kern-0.25em b}\kern-0.8em\TeX}}}

\copyrightyear{2020}
\acmYear{2020}
\setcopyright{acmlicensed}
\acmConference[AIES '20]{Proceedings of the 2020 AAAI/ACM Conference on AI, Ethics, and Society}{February 7--8, 2020}{New York, NY, USA}
\acmBooktitle{Proceedings of the 2020 AAAI/ACM Conference on AI, Ethics, and Society (AIES '20), February 7--8, 2020, New York, NY, USA}
\acmPrice{15.00}
\acmDOI{10.1145/3375627.3375815}
\acmISBN{978-1-4503-7110-0/20/02}

\begin{document}

\fancyhead{}
\title{The Offense-Defense Balance of Scientific Knowledge: Does Publishing AI Research Reduce Misuse?}


\author{Toby Shevlane}
\affiliation{%
  \institution{Faculty of Law}
  \institution{Centre for the Governance of AI, \linebreak Future of Humanity Institute}
\institution{University of Oxford}
}
\email{toby.shevlane@law.ox.ac.uk}

\author{Allan Dafoe}
\affiliation{%
  \institution{Department of Politics and \linebreak International Relations}
    \institution{Centre for the Governance of AI, \linebreak Future of Humanity Institute}
\institution{University of Oxford}
}
\email{allan.dafoe@governance.ai}


\begin{abstract}
There is growing concern over the potential misuse of artificial intelligence (AI) research. Publishing scientific research can facilitate misuse of the technology, but the research can also contribute to protections against misuse. This paper addresses the balance between these two effects. Our theoretical framework elucidates the factors governing whether the published research will be more useful for attackers or defenders, such as the possibility for adequate defensive measures, or the independent discovery of the knowledge outside of the scientific community. The balance will vary across scientific fields. However, we show that the existing conversation within AI has imported concepts and conclusions from prior debates within computer security over the disclosure of software vulnerabilities. While disclosure of software vulnerabilities often favours defence, this cannot be assumed for AI research. The AI research community should consider concepts and policies from a broad set of adjacent fields, and ultimately needs to craft policy well-suited to its particular challenges. 
\end{abstract}




\maketitle

\section{Introduction}

Although the vast majority of scientific research is socially beneficial, a subset of scientific fields struggle with the fact that their research can have harmful consequences. Scientists in these fields must weigh the potential harms in deciding what research to pursue, and whether, and how, to disclose any results. Broadly, approaches to disclosure can be arrayed on a spectrum between full publication and non-disclosure. Some fields, like computer security, are characterized by high levels of disclosure; researchers who discover software vulnerabilities tend to share them publicly, at least after initially warning vendors with the ability to patch them. Other domains, such as biosecurity, synthetic biology, cryptography, and nuclear security, exhibit greater secrecy and caution overall, with disclosure norms that are often actively discussed and contested.

The field of AI is in the midst of a discussion about its own disclosure norms, in light of the increasing realization of AI's potential for misuse. AI researchers and policymakers are now expressing growing concern about a range of potential misuses, including: facial recognition for targeting vulnerable populations, synthetic language and video that can be used to impersonate humans, algorithmic decision making that amplifies biases and unfairness, and drones that can be used to disrupt air-traffic or launch attacks \citep{brundage_malicious_2018}. If the underlying technology continues to become more powerful, additional avenues for harmful use will continue to emerge.

Some researchers argue that the field should move toward greater restraint in disclosing code and results, in cases where open publication would empower bad actors (see \citep{solaiman_release_2019}). Others argue that the field should be actively open in publishing such research, in order to increase awareness of risks and support actors who are working on defences \citep{zellers_why_2019}. In this paper we investigate this question of the \textit{security value of disclosure}: how publications and other forms of knowledge-sharing may increase or decrease the harms from misuse. We do not consider the broader consequences of scientific openness, most of which are beneficial.

The underlying tension is that a particular publication can be helpful both for individuals seeking to misuse the technology, but also for those seeking to prevent misuse. This paper provides a theoretical framework for thinking through this tension in a structured way. This will depend upon features of the social and technical dynamics surrounding the field, such as the cost of protecting against misuse, and the number of actors who are in a position to reproduce the relevant research. Much more policy work will be needed to identify best practice in measuring and weighing these factors.

Due to wide variation across fields, there is no reason to assume that disclosure norms that are appropriate for any given field will also be appropriate for AI. However, we show that the existing conversation around AI has heavily borrowed concepts and conclusions from one particular field: vulnerability disclosure in computer security. We caution against AI researchers treating these lessons as immediately applicable. There are important differences between vulnerabilities in software and the types of vulnerabilities exploited by AI. It is therefore important to explore analogies with multiple fields and to consider any properties that may make AI research unique. Ultimately, we suggest that the security benefits of openness are likely weaker within AI than in computer security; AI researchers are right to think hard about the implications of their research directions, their norms for sharing results, and the construction of institutions for mitigating downside risks.

The paper is organised as follows. First, we examine the recent debates around publication of language models; this case illustrates AI researchers' reliance upon ideas borrowed from computer security. Then, we provide our theoretical framework, which allows for a more structured comparison between fields. Finally, we apply the framework to AI research.

\section{Text Generation and the Computer Security Analogy}

Beginning in February 2019, a debate took place within the AI community about norms around the publication of potentially dangerous research results. This debate was provoked by OpenAI's public decision, and associated press, to not immediately share the trained weights of their larger ``GPT-2'' language models, because of concerns about the potential misuse of the ability to mass-produce plausibly human-written text, from fake product reviews, fake comments on news articles, to fake news itself \citep{radford_better_2019}. OpenAI adopted a policy of ``staged release'' of the GPT-2 model, referring to this as ``an experiment in responsible disclosure''. The term ``responsible disclosure'' originates from within computer security, and the staged release of software vulnerabilities (first to the software maker, and later to the public) has long been practised in computer security.

Criticism of OpenAI's approach has similarly been grounded in concepts and language borrowed from computer security. Indeed, the concepts imported from computer security are especially well-suited to arguing for prompt and full disclosure of potentially risky research (see below). Teams publishing generative language models after GPT-2 were particularly reliant on ideas from the vulnerability disclosure debates.

Grover, a model built by a team at the University of Washington, is an example. Grover was explicitly designed to generate disinformation or ``fake news'', under the belief that it is helpful to mimic how a malicious actor might use AI to cause harm. The research was published alongside code for building the model and dataset; for a period, researchers could email the authors to obtain the larger model weights, until they were open sourced several months later. The intention behind the research was to increase our understanding of the threat from text generation, and contribute to building tools that could guard against the harms. These benefits are supposed to outweigh any negative impacts of providing help to those who would use this technology for harm. The justification will be familiar to those well-versed in debates about the disclosure of software vulnerabilities: malicious actors will find it easy to build their own model regardless; knowledge of the attack is useful for preparing defences; and ``security through obscurity'' is ineffective.\footnote{Interview with Rowan Zellers; and \citet{zellers_why_2019}.}

A number of other researchers publishing similar language models have adopted this position: that the security benefits of publication outweigh the security risks. For example, Richard Socher, chief scientist at Salesforce, originally criticised OpenAI's delayed release of GPT-2 on the grounds that these language models are not dangerous. When later justifying full and immediate publication of his own team's model, \citet{socher_introducing_2019} stated that these models do carry dangers, but that such dangers weigh in favour of greater openness, with reference to computer security. He wrote: ``Similar to information security research, it is necessary for these tools to be accessible, so that researchers have the resources that expose and guard against potentially malicious use cases.''

Two master's students at Brown University replicated and published a version of GPT-2 \citep{gokaslan_opengpt-2:_2019}. In doing so, they sought to demonstrate that replication would be straightforward for a malicious actor, and therefore that partial publication is futile. During our interview, Aaron Gokaslan offered the example of software vulnerability disclosure, where the publication delay is often only around one month in duration, to critique the several-month delay adopted by OpenAI.

The imported arguments often make sense within a computer security context. For example, knowledge of a software vulnerability is very useful for developing a defensive fix (``patch''). However, from field to field, there will be variation in the extent to which assumptions like these hold. If AI research knowledge has a defensive bias, then this must be inferred through special attention to the relevant properties of AI research.

\section{Existing Literature: the Offense-Defense Balance}

The field of international relations has an existing body of literature on the offense-defense balance. The term refers to the ``relative ease of carrying out and defending against attacks'' \citep[p.738]{garfinkel_how_2019} \citep{jervis_cooperation_1978}. Scholars have studied the offense-defense balance of particular technologies such as the machine gun or cyber-weapons.\footnote{We can interpret this as the impact of these technologies on the underlying offense-defense balance within a conflict domain.} We extend the concept to apply to an item of knowledge: the relative benefit the knowledge offers to those attacking and defending within some existing context.

\citet{swire_model_2004}, a legal scholar, studies when disclosure improves security, and contrasts military affairs with the security of open source software. Swire argues that public knowledge of software code helps defenders more than hackers, because it is difficult to keep flaws in software secret, and because knowledge of flaws stimulates programmers to adapt the code. Conversely, militaries are unlikely to publish information about their defences (such as the location of land mines) because they have little to learn from outsiders.

\citet{shapiro_is_2010} analyse when states should publish information that may be useful to terrorists. For example, a government might wonder whether to publish weaknesses in commercial nuclear power plants. Like Swire, they highlight the tradeoff between aiding terrorists and stimulating defensive activity. Shapiro and Siegel find that the security effect of disclosure will depend on the likelihood that the attackers already hold the knowledge, and on the government's ability to leverage openness into finding and fixing new weaknesses.

\section{A framework for Weighing Security Costs and Benefits of Disclosure}

In this section we provide a framework for evaluating whether publication will promote or mitigate misuse. This framework could be used to make assessments specific to a field, to a subfield, or even to a single publication. However, the primary purpose in this paper is to highlight the relevant properties of AI research, as distinct from (or analogous to) other fields.

Our central question is whether disclosing research that may have harmful applications will, on balance, promote or mitigate those potential harms. The framework addresses the ``offence-defence balance of knowledge'', i.e. the balance of security costs and security benefits from disclosure.

As such, our analysis is limited in scope, only dealing with a subset of the consequences of disclosure. In terms of harms, we focus on direct, intentional use of the technology. We do not address cases where technology is used incautiously, or contains an unsafe defect, although in Appendix B we suggest that our analysis is relevant to such cases. Nor do we address structural risks from technological proliferation, such as military instability or unemployment.

Our framework does \textbf{not} weigh non-security benefits of publishing scientific results. Therefore, it should not be the basis for an assessment of publication norms, but rather one input to such an assessment. Potential ``non-security'' benefits of scientific publication include: contributions to economic growth and quality of life; advancement of science and its broad societal benefits; better monitoring of scientific investments; internationalism and cosmopolitanism from the global scientific community; and greater civilian control and involvement in science. Thus, even in cases where, under our framework, disclosure is considered as having negative security value (contributing more to harms than to their alleviation), it might nevertheless be the case that the general benefits of sharing the technology outweigh these harms.

\subsection{Theoretical Framework}

Disclosing research may make it easier for certain actors to cause harm. However, disclosure may also help certain actors to protect themselves or others from harm. The net effect of disclosure on misuse then depends on which of these two effects is stronger. Different types of disclosure will vary the effect, especially by changing what knowledge is shared, to whom, and when.

Let us first consider factors where disclosure affects the attacker's capacity to cause harm:
\begin{enumerate}
    \item \textbf{Counterfactual possession.} Counterfactual possession is where the would-be attacker acquires the relevant knowledge even without publication. Probable counterfactual possession will reduce the impact of the publication. Potential attackers have three main avenues for counterfactual possession:
    \begin{enumerate}
        \item \textit{Independent discovery}. In some cases, the attacker would be capable of producing or uncovering the knowledge themselves. This will depend on the distribution of expertise and resources across different populations. At one end of the spectrum, the relevant expertise and resources are exclusively held within communities that would not misuse it; at the other end of the spectrum, potential attackers have world-leading expertise and resources.
        \item \textit{Sharing amongst potential attackers}. How readily will knowledge spread between potential attackers? This factor amplifies independent discovery.
        \item \textit{Counterfactual publication.} Researchers sometimes argue that the risks from publication are not consequential, since the same research will soon be published by a different group \citep{leibowicz_when_2019}. A similar argument is that the publication only makes an incremental contribution, due to pre-existing literature. However, we believe these considerations should be excluded from the decision. Instead of considering the impact of an individual decision to publish, researchers should ask: ``what decision would I want rolled out across the whole field?''
    \end{enumerate}
    \item \textbf{Absorption and application capacity.} Publication of the research will only benefit attackers to the extent that they are able to absorb and apply the research. This will depend on a number of factors, such as:
    \begin{enumerate}
        \item \textit{Receptiveness to scientific research}. Will the research be read and understood by potential attackers? This will depend on the attackers' attentiveness and comprehension, but it will also depend on the disclosure itself: how much knowledge is disclosed, through what channel, and how is it presented?
        \item \textit{Sufficiency.} In extreme cases, the published research will contain all that is needed to carry out the harmful behaviour. On the other end of the spectrum, adopting the research will involve a large investment in resources and complementary knowledge \citep{lewis_information_2018}.
        \item \textit{Transferability.} Defensive knowledge (e.g. the design of a detection system) can sometimes be converted into a more effective attack (e.g. avoiding the detection system). The transferability of knowledge (from defensive to offensive, or vice versa) will vary across cases.
    \end{enumerate}
\end{enumerate}
Given the factors listed, we note that the actors benefiting most from publication will exist in a ``Goldilocks zone'' of capability, where they are sufficiently capable to apply the knowledge but not to independently discover it (see Appendix A for further detail).

We can also consider how disclosure will affect defenders' ability to mitigate potential harms. The disclosure could disseminate ideas that are useful for finding solutions, or it could sound the alarm. The success of these attempts to promote defences will be governed by a number factors:
\begin{enumerate}
    \item \textbf{Counterfactual possession.} See above. Would the defenders independently discover, or obtain through some other means, the knowledge contained in the publication?
    \begin{enumerate}
        \item Where the proposed publication offers some defensive insight, how easily could the defenders have arrived at that insight, or a similarly good one, themselves?
        \item Where the publication contributes to defences by sounding the alarm, would the defenders have already been aware of the problem? In some cases, there will be many hidden vulnerabilities to attack, and the knowledge of a particular vulnerability will be high-value. In other cases, there will only be a single, obvious way to misuse the technology, and so sounding the alarm will be redundant.
    \end{enumerate}
    \item \textbf{Absorption and application capacity.} See above.
    \item \textbf{Resources for solution-finding.} Given the disclosure, how many additional individuals or organisations will work on finding a solution?
    \item \textbf{Availability of effective solutions.} The positive effects of disclosure depend on the potential for a good defence against the misuse. In some cases, an effective solution is possible, but in other cases, no effective solution exists.
    \begin{enumerate}
        \item Is the weakness baked into the fundamentals of the system, or could a relatively superficial change remove it?
        \item Is the attack detectable, and is detection sufficient for defence or deterrence?
        \item Is the attack so powerful that it will overwhelm any defences? I.e. how offense-dominant is the technology itself?
    \end{enumerate}
    \item \textbf{Difficulty/cost of propagating solution.} Even where a solution exists in theory, it might be difficult or costly to propagate that solution.
    \begin{enumerate}
        \item \textit{Centralisation. }In some cases a solution can be imposed centrally, such as a change to the code of a website, whereas in other cases, the solution requires action from a large population of individuals.
        \item \textit{Solution adoption.} Solutions vary in how readily they will be adopted. For example, it might be relatively easy for users to update their software, but it might be costly for individuals to change physical hardware, such as locks, or their social arrangements.
    \end{enumerate}
\end{enumerate}

The offense-defense balance of a given disclosure strategy will depend on weighing these benefits against the extent to which disclosure contributes to misuse. For misuse risks that have a higher potential harm (e.g. a vulnerability in Facebook versus a vulnerability in a much smaller website), the security consequences of disclosure (positive or negative) will be amplified accordingly.

Assessment should be done across the range of available disclosure strategies, allowing for the security value of different forms of disclosure to be compared. For example, in principle AI researchers can choose to publish more or less:  basic results and insights, detailed results, code, datasets, trained networks, and easy-to-use tools. These different outputs will differentially benefit different actors (see Appendix A). A publication without practical tools or code will be more difficult for low-capability attackers (and defenders) to apply; although even here, the publication will often trigger further work that lowers the barriers to application. Also, researchers could attempt to give their release a defensive bias by investing extra effort in preparing defensive tools and best-practice, and differentially publishing those. Researchers can also circulate certain knowledge or tools exclusively among the scientific community.

\section{Contrasting Different Fields}

The framework helps to explain why the disclosure of software vulnerabilities will often be beneficial for security. Patches to software are often easy to create, and can often be made in a matter of weeks. These patches fully resolve the vulnerability. The patch can be easily propagated: for downloaded software, the software is often automatically updated over the internet; for websites, the fix can take effect immediately. In addition, counterfactual possession is likely, because it is normally easier to find a software vulnerability (of which there is a constant supply) than to make a scientific discovery (see \citep{bloom_are_2017}). These factors combine to make a reasonable argument in favour of public disclosure of software vulnerabilities, at least after the vendor has been given time to prepare a patch.

Contrasting other fields will further bring into relief the comparatively defence-dominant character of software vulnerability knowledge. We can focus on the tractability of defensive solutions: for certain technologies, there is no low-cost, straightforward, effective defence.

First, consider biological research that provides insight into the manufacture of pathogens, such as a novel virus. A subset of viruses are very difficult to vaccinate for (there is still no vaccination for HIV) or otherwise prepare against. This lowers the defensive benefit of publication, by blocking a main causal pathway by which publication leads to greater protection. This contrasts with the case where an effective treatment can be developed within a reasonable time period, which could weigh in favour of publication \citep{lewis_information_2018}.

Second, consider cases of hardware based vulnerabilities, such as with kinetic attacks or physical key security. Advances in drone hardware have enabled the disruption of airports and attacks on infrastructure such as oil facilities; these attacks presently lack a cheap, effective solution \citep{rawlinson_heathrow_2019}. This arises in part from the large attack surface of physical infrastructure: the drone's destination can be one of many possible points on the facility, and it can arrive there via a multitude of different trajectories. This means that the path of the drone cannot simply be blocked.

Moreover, in 2003 a researcher published details about a vulnerability in physical key systems \citep{blaze_rights_2003}. Apartment buildings, offices, hotels and other large buildings often use systems where a single master-key can open all doors. The research showed how to derive the master-key from a single non-master key. The researcher wrote that there was ``no simple or completely effective countermeasure that prevents exploitation of this vulnerability short of replacing a master keyed system with a non-mastered one'' (\citep{blaze_master-keyed_2003}; see \citep{blaze_rights_2003} for further discussion of counter-measures). The replacement of master-key systems is a costly solution insofar as master-key systems are useful, and changes are very difficult to propagate: physical key systems distributed across the world would need to be manually updated.

Finally, consider the policy question of whether one should have published nuclear engineering research, such as on uranium enrichment, in the 1960s. For countries like India and Pakistan, this would have increased, not decreased, their potential to destroy each others' cities, due to the lack of defensive solutions: as with certain diseases, nuclear bombs cannot be adequately protected against. Moreover, for the minor protections against nuclear bombs that exist, these can be pursued without intricate knowledge as to how nuclear bombs are manufactured: there is low transferability of offensive into defensive knowledge.\footnote{Although basic knowledge of the bomb would be useful, such as the size of explosions, or the threat of radioactive fallout.} For example, a blueprint for the design of a centrifuge does not help one build a better defensive bunker. Overall, if both a potential defender and potential attacker are given knowledge that helps them build nuclear weapons, that knowledge is more useful for making an attack than protecting against an attack: the knowledge is offense-biased.\footnote{However, note that the proliferation of nuclear weapons can have a deterrence effect: attackers become vulnerable to retaliation.}

Differences across fields will shape the security value of publication, which can influence disclosure norms among security-minded scientists and policymakers. The Manhattan Project was more secretive than locksmiths and influenza researchers, who are in turn often more secretive than those finding vulnerabilities in software. Indeed, there was a culture clash between the researcher who published the flaw in the master-key system, above, who came from a computer security background, and the locksmiths who accused him of being irresponsible. The different disclosure cultures exist in the form of default practices, but also in common refrains - for example, language about the virtues of ``studying''a problem, or the value of users being empowered by disclosure to ``make decisions for themselves''. Such language embeds implicit answers to the framework given in this section, and therefore caution should be exercised when importing concepts and language from other fields.

\section{Applying the Framework to AI}

What is the offense-defense balance of knowledge within AI research? We focus on the general properties of AI research, despite likely variation within the field. (Appendix C discusses the appropriate level of generality at which AI projects should be assessed.)

Our analysis is preliminary and non-comprehensive. Our primary argument is that, relative to vulnerability discovery in computer security, many forms of AI misuse are more difficult and costly to defend against, because AI misuse often interferes in (complex, distributed, sticky) social systems. In general, this problem is not adequately solved by the possibility of detecting AI activity. Finally, we consider which actors (attackers or defenders) would be best equipped to independently discover the knowledge contained in the publication. Overall, we tentatively argue that within AI, the security value of publication will be net negative in a significant fraction of cases, and more so than in computer security. Further research on these points would be worthwhile.

\subsection{The Difficulty of Patching Social Vulnerabilities}

For a disclosure to have defensive benefit, protections against misuse must be tractable: solutions must be possible (point 4, above) and not prohibitively costly or difficult to impose (point 5). The tractability of solutions will depend on the domain (digital, physical, or social) that is being exploited. Digital systems are in general the easiest to patch at scale, whereas physical and social systems are often much more costly to change, especially at scale. To the extent that protections against AI misuse more often require interventions in social systems, as they seem to, then, all else equal, AI publications will have a more limited defensive benefit.

AI is especially prone to interfering in social systems because AI involves automating activity that normally requires human intelligence. AI systems may be able to mimic the sound of the human voice, or the way that humans construct sentences, or the way that humans play poker. This means that AI systems could step into the shoes of humans on platforms that host human interaction, such as social media, telecommunications, email, online banking, stock trading, commerce, or online gambling. This property of AI systems means that it has the potential to interfere with social relations, rather than relying on interference with computer or physical systems. The ``vulnerability'' may be an ordinary way of organising social life, which could therefore be very costly or difficult to change.

For example, consider a recent case where criminals used artificial speech generation to fake the voice of the CEO of a company over the telephone \citep{statt_thieves_2019}. This attack exploits the fact that people use the sound of someone's voice as strong evidence of their identity. Thus, the ``vulnerability'' is socially useful, and is deeply ingrained in human affairs. In computer security, it might be easy to propagate a patch to a line of code, but often it is not so easy or desirable to ``patch'' social systems. Indeed, even within computer security, it is often remarked that human behaviour is the weakest link (for example, it is difficult to encourage humans to use safe passwords).

The same problem has arisen in relation to AI that generates human-like text. Some have responded to concerns over text generation by suggesting that the research community should simply ``warn society'' that individuals may be increasingly shown fake or untrustworthy content. However, adjusting for the increasing untrustworthiness of one's informational environment could be burdensome, especially if it simultaneously leads to distrust of true information \citep{kreps_not_2019}. The language of ``let the users decide for themselves'', reminiscent of computer security discourse, would lose its empowering sentiment if users become landed with problems for which no good solution exists.

The tractability of solutions will vary from case to case, and in many cases, some solution (technical, institutional, or behavioural) will be able to counter AI misuses. The key suggestion is that the vulnerabilities will be on average harder to patch than software vulnerabilities, and so AI researchers cannot assume that their publication will always have a defensive bias.

\subsection{Detection of AI Misuse}

A common response to concerns over AI misuse is that there is defensive value in \textit{detecting} these AI attacks, using automated systems that can discern AI from human activity. An AI model that is useful for an attack could simultaneously be useful for detecting these attacks (e.g. \citep{zellers_defending_2019}). In such cases, this provides a possible defensive benefit to publishing the model.

Nonetheless, in such cases, the defensive benefit of disclosure will depend on the effectiveness and practical feasibility of these detection systems. In this regard, we highlight two potential difficulties.

First, offensive AI systems can often be trained against detection systems, so as to evade them, similar to how a biological disease evolves so as to be difficult to detect by the immune system. The team behind Grover, the AI model that generates artificial news articles, found that when Grover's generations were pre-filtered by the detection system, the remaining generations were harder to detect (unless the detection system was given many of these new generations to train on, like a vaccine) \citep{zellers_counteracting_2019}. This is one reason against making the detection system freely available.

Second, even in cases where a detection system can in theory detect the AI's activity, it may not be feasible to deploy this system. For example, consider the case of the voice impersonated CEO. Even if a system could detect when a voice was AI-generated, deploying this system might require surveillance of calls, large amounts of computation, and false-positive flagging of ordinary conversations.\footnote{Analogously, it is difficult to eliminate false positives for text-based detection systems: see \citep{gehrmann_gltr:_2019}.} These systems may also be financially expensive to set up, which could lead to non-deployment. In sum, the security value of publication, if reliant on improving detection systems, will depend on the practical limits of those detection systems.

\subsection{Counterfactual Possession of AI Knowledge}

The effect of publishing research will depend on \textit{counterfactual possession: }the extent to which attackers and defenders would have obtained the same knowledge regardless of publication. This depends on the extent to which scientific and engineering knowledge about AI, and resources for building AI systems (including computation), are concentrated in the mainstream AI research field, versus extending outward to attackers or defenders.

The issue of counterfactual possession has been raised within the debate over text generation. Where the risks of text generation come from state-led disinformation campaigns, one uncertainty is how much these state actors would benefit from the scientific research: do they employ researchers capable of independently conducting the research? This is certainly possible, although it does not fully eliminate the security costs of disclosure. Firstly, the population of wrongdoers should be looked at in its totality, rather than cherry-picking the most capable actors. State actors may be the most difficult to impact, but there will be many actors with less access to AI talent and compute. Furthermore, counterfactual possession will depend on the content of the disclosure. Although a range of actors may be capable of using a research paper to recreate the code or model, there may be very few actors outside the AI research community capable of having the original insights contained in the paper.

In addition, we must equivalently consider the defenders' ability to independently produce the knowledge. The identity of the defenders will vary from case to case. However, in some cases, it will be large technology companies who need to prepare defences. In the case of AI operating over digital platforms such as Facebook, it is worth noting that these companies have large expertise and resources: Facebook, for example, employs many AI researchers, and invests heavily in finding technical solutions to misuse of the platform \citep{roettgers_mark_2019}. In such cases, there is strong reason to believe that the defenders would be the actors most likely to independently discover the relevant research insight.

Taken together, the different arguments made in this section suggest that, where AI research carries misuse risks, researchers should be alert to the possibility that those risks will be exacerbated by publication. At the least, the security value of publication appears lower in AI research than in computer security.

\section{Conclusion}

Our analysis should aid AI researchers in thinking critically about the implications of publishing work that could be misused. As the discussions around AI misuse mature, the community should grow its toolbox of analogies and concepts, such that disclosure norms can be well-designed.

In the search for the right disclosure norms, our analysis is only one piece of the puzzle. The security impact of disclosure is important, but should be considered alongside a host of other considerations. One challenge will be building disclosure policies in accordance with a legitimate and effective norm-setting process \citep{crootof_artificial_2019}. Another issue will be how to maintain the benefits of open research. For example, openness currently allows for some transparency over the research projects taking place within labs, which would be necessary for holding labs accountable for the ethics and safety of their projects (c.f. \citep[p.8]{bostrom_strategic_2017}). Finally, it is important that the AI research community avoids overfitting to the present-day technologies, and finds norms that can scale and readily adapt to more powerful AI capabilities.

At the same time, disclosure norms are not the only tool for tackling harmful applications of AI. The research community can differentially invest in those projects and trajectories that are most socially beneficial. Researchers can invest extra effort in understanding and mitigating the potential harmful uses of their research, including through the crafting of norms and policies to steer the use of AI.

\begin{acks}

We are grateful for helpful comments and suggestions from: Markus Anderljung, Miles Brundage, Michael Cafarella, Rosie Campbell, Peter Cihon, Jack Clark, Ben Day, Jeff Ding, Jade Leung, Gregory Lewis, Luke Muehlhauser, Carina Prunkl, Jacob Shapiro, Carl Shulman, David Siegel, Steven Weber, Toby Ord, and especially Ben Garfinkel. We thank the Open Philanthropy Project for funding support.

\end{acks}

\bibliographystyle{ACM-Reference-Format}
\bibliography{sample-base}

\appendix

\section{Goldilocks Zone}

We have theorised that certain actors will benefit from scientific publication more than others. The actors on whom publication will have the greatest impact are those with sufficient capability to apply the knowledge, but without sufficient expertise to independently discover the knowledge. We can refer to this as the ``Goldilocks zone'' of capability. The following diagram shows the Goldilocks zone:
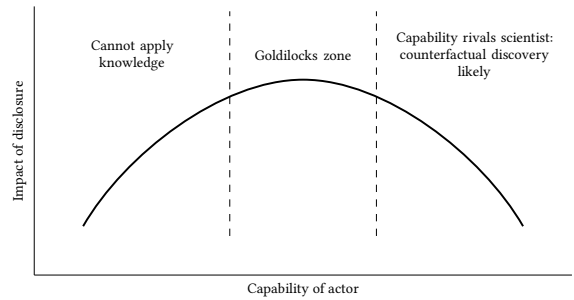
\begin{center}
\resizebox{0.9\linewidth}{!}{\begin{tikzpicture}[x=20pt,y=20pt]
    \draw (0,0)--(11,0) node[midway,below] {\scriptsize Capability of actor};
    \draw (0,0)--(0,5.5) node[midway,above,rotate=90] {\scriptsize Impact of disclosure};
    \draw[thick] (1,1) to[out=60,in=180,looseness=0.8] (5.5,4) to[out=0,in=120,looseness=0.8] (10,1);
    \begin{scope}[dashed]
        \draw (4,0.8)--++(90:4.6);
        \draw (7,0.8)--++(90:4.6);
    \end{scope}
    \begin{scope}[every node/.style={font=\scriptsize,text width=80pt,align=center}]
        \node at (2,4.5) {Cannot apply\\ knowledge};
        \node at (5.5,4.5) {Goldilocks zone};
        \node at (9,4.5) {Capability rivals scientist:\\ counterfactual discovery\\ likely};
    \end{scope}
\end{tikzpicture}}
\end{center}
The type of knowledge being disclosed will affect both the ability of actors to apply the knowledge, and the likelihood of independent discovery. Therefore, the slope of the curve will depend on the knowledge being disclosed.

For example, we can consider the extreme case of scientific knowledge that could only be produced by a handful of the world's most capable scientists. We can call this the ``Einstein case''. In this case, the effect of independent discovery will be removed, and the Goldilocks zone extended accordingly:
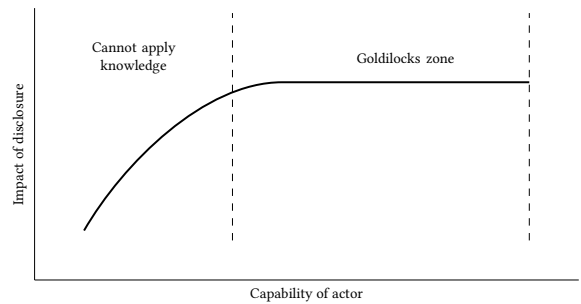
\begin{center}
\resizebox{0.9\linewidth}{!}{\begin{tikzpicture}[x=20pt,y=20pt]
    \draw (0,0)--(11,0) node[midway,below] {\scriptsize Capability of actor};
    \draw (0,0)--(0,5.5) node[midway,above,rotate=90] {\scriptsize Impact of disclosure};
    \draw[thick] (1,1) to[out=60,in=180,looseness=0.8] (5,4) -- (10,4);
    \begin{scope}[dashed]
        \draw[dashed] (4,0.8)--++(90:4.6);
        \draw[dashed] (10,0.8)--++(90:4.6);
    \end{scope}
    \begin{scope}[every node/.style={font=\scriptsize,text width=80pt,align=center}]
        \node at (2,4.5) {Cannot apply\\ knowledge};
        \node at (7.5,4.5) {Goldilocks zone};
    \end{scope}
\end{tikzpicture}}
\end{center}
We could also consider the case of knowledge that requires very little capability to apply. For example, with the right exploit code, some software vulnerabilities are very easy to exploit, requiring very little expertise. We can call this the ``script kiddie'' case. The Goldilocks zone is extended, in this case to the left:
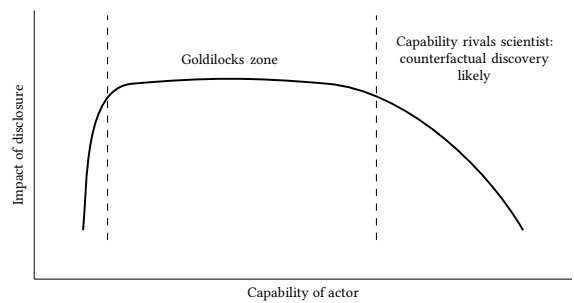
\begin{center}
\resizebox{0.9\linewidth}{!}{\begin{tikzpicture}[x=20pt,y=20pt]
    \draw (0,0)--(11,0) node[midway,below] {\scriptsize Capability of actor};
    \draw (0,0)--(0,5.5) node[midway,above,rotate=90] {\scriptsize Impact of disclosure};
    \draw[thick] (1,1) to[out=85,in=185,looseness=0.8] (2,4) to[out=5,in=175] (6,4)
     to[out=355,in=120,looseness=0.8] (10,1);
    \begin{scope}[dashed]
        \draw[dashed] (1.5,0.8)--++(90:4.6);
        \draw[dashed] (7,0.8)--++(90:4.6);
    \end{scope}
    \begin{scope}[every node/.style={font=\scriptsize,text width=80pt,align=center}]
        \node at (4,4.5) {Goldilocks zone};
        \node at (9,4.5) {Capability rivals scientist:\\ counterfactual discovery\\ likely};
    \end{scope}
\end{tikzpicture}}
\end{center}
AI research publications often contain various forms of knowledge (e.g. scientific insight; code, model weights), which would independently each have different Goldilocks zones. In other words, different elements of the publication are more impactful for different actors. If the scientific insight is strong, then it would be especially impactful for those high-capability actors who could not quite achieve that level of independent discovery. The code will be especially useful for lower capability actors, because the high capability actors could reverse engineer the code (assuming the scientific paper is published). The model weights will be especially useful for actors without the financial resources to train the model, depending on the compute-intensity of the training process.

\section{Generalising to Unintended Harms}

We have been referring to ``attacks'' and ``attackers''. However, an analogous problem occurs where agents use scientific research in a way that generates unintended consequences. Many harms from technology are unintended accidents, byproducts, or emergent consequences \citep{bostrom_information_2011}. For example, some biology labs handling diseases might have poor biosafety standards, risking an outbreak. Another example is the graduate student who intends to spread a harmless computer worm, but the worm turns out to be much more virulent than he expected (see the ``Morris worm''). Similar examples have arisen within machine learning: it is commonly claimed that the YouTube recommender system favours politically radical content, despite no such intention from the designers. One could also imagine a text generation system that (due to artefacts of the dataset or training process) was prone to politically controversial, racist or sexist, or otherwise offensive generations.

Researchers may want to take account of unintended harms when making publication decisions. Although this is not the main focus of our paper, we suggest that our framework could be relevant for considering unintended harms. Instead of ``attackers'', there would be agents who use the technology in a negligent way (unless the defect is so strong that even cautious users will fall victim) or deprioritise the risks in favour of other goals. Under the framework, the researcher would assess whether these use-cases will be furthered by publication, by looking at the extent to which publication will increase the agents' liability toward using the technology in a harm-causing way (depending on their absorption and application capacity, and counterfactual possession). The defensive considerations will also remain: in publishing, the researcher might hope to draw attention to the risks, or further the development of solutions.

Nevertheless, there may be new considerations raised by the case of unintended consequences. One important difference is the location to which defences can be applied. In the case of unintended consequences, the dangers will be posed by a deficiency of the technology (rather than a malicious application). Therefore, a key solution will be fixing the technology itself (rather than just the environment in which it operates). This could either be done post-publication by other researchers, or pre-publication. For example, in the case of the profane text generator, publication could be delayed until researchers have solved the problem.

\balance

\section{Assessing Strands of Research Rather than Individual Papers}

When making disclosure decisions, an important question will be whether decision-makers should focus narrowly on the research paper at hand, or consider the more general method or technology that is being contributed to.

On the one hand, the category of ``AI'' is broad, subsuming specific technologies with different security costs and benefits of disclosure. It might be that knowledge of how to build a lie-detection AI has a different offence/defence profile from that of a robotic soldier, which in turn is different from an AI that can pretend to be a human on a dating app. Therefore, the AI community should be ready to consider a particular strand of research on its merits.

At the same time, we would argue that researchers should not narrowly focus on the details of a single paper. The risk is that researchers fail to foresee cases where their research contributes to a future misuse only once it has been further improved, or applied to a new problem. For example, research on image generation can subsequently become useful for stitching celebrity faces onto pornographic videos. Also, when OpenAI only partially published its GPT-2 model, one objection was that GPT-2-produced text was difficult to direct, and thus the research would not be useful for automating disinformation campaigns. However, subsequent research has already begun to resolve this weakness.\footnote{CTRL and Grover both take in metadata, such as the title of an article: \citep{keskar_ctrl:_2019}, \citep{zellers_defending_2019}.} In this context, we would recommend that AI research papers should be assessed in terms of the subsequent family of models and methods that will emerge, rather than considering the weaknesses of a single model in isolation from the longer-term research direction \citep{ovadya_reducing_2019}.

\end{document}